\documentclass[12pt]{article}
\usepackage{amsfonts, amsmath, amssymb, cite, graphicx, }

\usepackage{color}
\usepackage[all, knot]{xy}
\usepackage{tikz}
\usepackage{subfigure}
\usepackage{braket}

\usepackage[utf8]{inputenc}
\usepackage{ulem}

\usepackage{bm}
\usepackage{epstopdf}
\usepackage[footnotesize]{caption}
\usepackage{amsthm}
\usepackage{amsthm,amsmath,amssymb}
\usepackage{enumitem}
\usepackage{mathrsfs}
\usepackage{makecell}
\usepackage{diagbox}

\usepackage[margin=3cm]{geometry}

\begin{document}

\begin{titlepage}
%\begin{flushright}
%TIT/HEP-6XX \\
%mm,  2016
%\end{flushright}
\vspace{0.5cm}
\begin{center}
{\Large \bf A possible connection between quantum mechanics and spacetime}

\lineskip .75em
\vskip 2.5cm
{\large Hong Wang$^{a}$, Jin Wang$^{b,c,}$\footnote{jin.wang.1@stonybrook.edu} }
\vskip 2.5em
 {\normalsize\it $^{a}$State Key Laboratory of Electroanalytical Chemistry, Changchun Institute of Applied Chemistry, Chinese Academy of Sciences, Changchun 130022, China\\
 $^{b}$Center of Theoretical Physics, College of Physics, Jilin University, Changchun 130012, China\\
 $^{c}$Department of Chemistry and Department of Physics and Astronomy, State University of New York at Stony Brook, NY 11794, USA}
\vskip 3.0em
\end{center}
\begin{abstract}
Recent developments in holographic gravity suggest that spacetime structure may be deeply related to quantum mechanics. In this work, from a different perspective, we demonstrate that wave--particle duality can be interpreted as the uncertainty of spacetime for the particle. Summarizing all possible trajectories in conventional path integral quantum mechanics can be transformed into the summation of all possible spacetime metrics. Furthermore, we emphasize that in conventional quantum gravity,  it is possible that the classical matter fields correspond to quantum spacetime. We argue that this is not quite reasonable and propose a new path integral quantum gravity model based on the new interpretation of wave--particle duality. In this model, the aforementioned drawback of conventional quantum gravity naturally disappears.
\end{abstract}
\end{titlepage}

\baselineskip=0.7cm

\tableofcontents
\newpage
\section{Introduction}

\label{sec:0}
Quantum mechanics has been studied for nearly a century. Despite ongoing debates regarding the interpretation of the wave function \cite{HE,GL23,GW23,CR23,MB23,SG23,AH23,SW231,SW232,SW233,AK23,GAT23,JG23,RB23,WH23,JP910,AC910,LV910,AN910,CT910}, it has become one of the greatest theories. Quantum theories for the weak, strong and electromagnetic interactions have been successfully developed. These theories collectively constitute the renowned Standard Model of particle physics. However, a self-consistent quantum theory of gravity is still lacking \cite{CK,CR}. Therefore, it is significant to make new attempts to construct a theory of quantum gravity.

There are two equivalent methods for quantizing a system: the path integral and canonical quantization. In path integral quantum mechanics~\cite{RP910},  to calculate the transition amplitude, one must sum up all possible trajectories. The trajectory with the minimum action is the classical trajectory. For a free particle, it is a straight line. Other trajectories are not straight lines and do not satisfy the Hamilton--Jacobi equation. Only the classical trajectory  corresponds clearly to a dynamical Equation (Hamilton--Jacobi equation). Thus, the classical trajectory possesses a special status. However, it may be  more natural to consider all trajectories solutions of some dynamical equations. What is this relevant dynamical equation? Is this idea feasible, and what implications does it have for quantum~mechanics?

In this work, we show that different trajectories in the path integral can be regarded as solutions of the Hamilton--Jacobi equation in different spacetimes. For the free particle, different trajectories in the path integral can be interpreted as the geodesics in different spacetimes. We find that the integration over all possible trajectories in the path integral quantum mechanics can be transformed into the integration over various spacetime metrics. This results in the spacetime representation of quantum mechanics. We show that in the spacetime representation,  when using the path integral to quantize any system, one only needs to set the spacetime metric as the integral variable.  We demonstrate that wave--particle duality can be interpreted as the uncertainty of the spacetime structure for the microscopic particle. Thus, the quantum characteristics of any system can be attributed to the uncertainty of the spacetime structure. This result suggests that quantum mechanics is deeply connected to spacetime structure. Recent developments in holographic gravity also suggest a deep relationship between quantum mechanics and spacetime~\cite{ST,MV211,BS211,YN211,BM211}.

In conventional path integral quantum gravity, classical matter fields may correspond to the quantum spacetime~\cite{HJ}. We demonstrate that this is not quite reasonable. Based on the new interpretation of wave--particle duality, we propose that in path integral quantum gravity, only the spacetime metric should be considered the integral variable.  By adopting this proposal, the aforementioned problem in quantum gravity can be easily circumvented. Throughout this study, we work in units where $G=\hbar=c=1$.

%%%%%%%%%%%%%%%%%%%%%%%%%%%%%%%%%%%%%%%%%%%%%%%%%%%%%%%%%
%%%%%%%%%%%%%%%%%%%%%%%%%%%%%%%%%%%%%%%%%%%%%%%%%%%%%%%%%
%%%%%%%%%%%%%%%%%%%%%%%%%%%%%%%%%%%%%%%%%%%%%%%%%%%%%%%%%
%%%%%%%%%%%%%%%%%%%%%%%%%%%%%%%%%%%%%%%%%%%%%%%%%%%%%%%%%

\section{Spacetime Representation of Quantum Mechanics}
\label{sec:1}

\subsection{A Single Particle}
\label{sec:1.1}
In quantum mechanics,  there are still conceptual puzzles regarding wave--particle duality and the wave function.  The evolution of the wave function is governed by the Schr\"{o}dinger equation, which can be expressed in various representations. Different representations are  connected through  a unitary transformation. In this section, we will explore the possibility of  a spacetime representation in quantum mechanics, which may help to understand wave--particle duality and the wave function.

The spacetime coordinates can be chosen arbitrarily. Cartesian coordinates are convenient for our illustration. We will use symbols $X^{\mu}$,  $X^{\alpha}$, $X^{\beta}$, \ldots to represent Cartesian coordinates. Other coordinates will be denoted by different symbols, such as  $x^{\mu}$, $(\theta, \varphi)$, \ldots. 
{We denote the spacetime metric as $g_{\alpha\beta}$}.  When the spacetime is flat, Cartesian coordinates imply that $g_{\alpha\beta}=\eta_{\alpha\beta}$, where $\eta_{\alpha\beta}=diag(1,-1,-1,-1)$ represents the Minkowski metric.

For simplicity, we start by examining a relativistic free particle in a curved spacetime. The classical  dynamics of the particle are determined by its action,  expressed as~\cite{E}
\begin{equation}
\label{eq:1.1}% 1
S=m_{0}\int_{\tau_{i}}^{\tau_{f}}d\tau\sqrt{g_{\alpha\beta}\frac{dX^{\alpha}}{d\tau}\frac{dX^{\beta}}{d\tau}},
\end{equation}
{where $-m_{0}$  represents the rest mass of the particle. The negative sign is explicitly factored out to facilitate  the discussion of dynamics in the non-relativistic limit. } $\tau$ is the proper time parameter.  Taking $\delta S/\delta X^{\mu}=0$, one can obtain the classical dynamical Equation (geodesic equation) of the particle as
\begin{equation}
\label{eq:1.2}% 2
\frac{d^{2}X^{\mu}}{d\tau^{2}}+\Gamma^{\mu}_{\alpha\beta}\frac{dX^{\alpha}}{d\tau}\frac{dX^{\beta}}{d\tau}=0.
\end{equation}
Here, $\Gamma^{\mu}_{\alpha\beta}$ is the connection.

In a specific spacetime $g_{\alpha\beta}$, if the initial state of the particle is fixed, then Equation~\eqref{eq:1.2} has a unique solution. Changing the metric $g_{\alpha\beta}$ continuously, the solution may also be changed continuously. Thus, the trajectory of the particle can be seen as a functional of the metric.  We formally denote the solution to the geodesic Equation \eqref{eq:1.2} as
\begin{equation}
\label{eq:1.3}% 3
X^{\mu}(\tau)=\mathcal{F}^{\mu}(g_{\alpha\beta}).
\end{equation}
Equation \eqref{eq:1.2} is a nonlinear equation. It is difficult to give a general solution for the geodesic equation. Thus, the explicit form of the functional $\mathcal{F}^{\mu}$ is difficult to specify. However, the conclusions of this work are independent of the detailed form of the functional $\mathcal{F}^{\mu}$. We point out that the metric $g_{\alpha\beta}$ and the solution to Equation \eqref{eq:1.2} are not in a one-to-one correspondence. Fixing the metric $g_{\alpha\beta}$ and the initial state of the particle, there is only one trajectory that satisfies Equation \eqref{eq:1.2}. Conversely, given the trajectory and the initial state of the particle, there are infinite spacetime metrics that are compatible with the trajectory of the particle. In other words, one can not determine the spacetime structure (metric)  according to the trajectory of the free particle.

Equation \eqref{eq:1.2} can also be interpreted as a particle moving in a gravitational field. When the gravitational field strength is zero, the trajectory of the particle is a straight line. By changing the gravitational field, the trajectory can take on any curve. Thus, it is reasonable to infer that for any given curve (not violating physical principles), there exists a gravitational field that causes it to become the particle's trajectory. In other words, given any one physically allowed curve, it can always be embedded into a spacetime where it becomes a geodesic.  By changing the local spacetime structure far from the trajectory of the particle, the dynamics of the particle remain unaffected. Hence, for any physically allowed curve, there are infinite types of spacetime where one of the geodesics matches this curve.

In Minkowski spacetime,  according to conventional path integral quantum mechanics, the transition amplitude of a relativistic particle  from the initial state $|X^{\mu}(\tau_{i})\rangle$ to the final state $|X^{\mu}(\tau_{f})\rangle$ can be written as\vspace{-6pt}
\begin{equation}
\label{eq:1.4}% 4
 \langle X^{\mu}(\tau_{f})|X^{\mu}(\tau_{i})\rangle=N\int DX^{\mu}(\tau)\ \mathrm{exp}\{i m_{0}\int_{\tau_{i}}^{\tau_{f}}d\tau\sqrt{\eta_{\alpha\beta}\frac{dX^{\alpha}}{d\tau}\frac{dX^{\beta}}{d\tau}}\}.
\end{equation}
Here, $N$ represents the normalized constant.  $\int DX^{\mu}(\tau)$ represents the summation over all trajectories connecting the initial and final states. The factor $\int_{\tau_{i}}^{\tau_{f}}d\tau\sqrt{\eta_{\alpha\beta}\frac{dX^{\alpha}}{d\tau}\frac{dX^{\beta}}{d\tau}}$ represents the length of the trajectory $X^{\mu}(\tau)$ between the initial and final state. The lengths of different trajectories may be different.  The classical trajectory is the solution to the Hamilton--Jacobi equation, while  all other trajectories do not satisfy the classical dynamical equation. According to the conventional interpretation, the spacetime background in the path integral \eqref{eq:1.4} is deterministic (Minkowski spacetime). Various trajectories in Equation \eqref{eq:1.4} are curves within Minkowski spacetime.

Every trajectory in the path integral \eqref{eq:1.4} is a possible path of the particle from the initial state $|X^{\mu}(\tau_{i})\rangle$ to the final state $|X^{\mu}(\tau_{f})\rangle$. Furthermore, each trajectory  should be physically allowed. As discussed previously, for any given trajectory $X^{\mu}(\tau)$ in the path integral \eqref{eq:1.4}, there exists a spacetime where this trajectory is a geodesic. (This point is crucial for this study. We provide a more rigorous demonstration of this point in  Appendix~\ref{sec:A}.) Therefore, the length of the trajectory can be written as
\begin{equation}
\label{eq:1.5}% 5
\int_{\tau_{i}}^{\tau_{f}}d\tau\sqrt{\eta_{\alpha\beta}\frac{dX^{\alpha}}{d\tau}\frac{dX^{\beta}}{d\tau}}=\int_{\tau_{i}}^{\tau_{f}}d\tau\sqrt{g_{\alpha\beta}\frac{dX^{\alpha}}{d\tau}\frac{dX^{\beta}}{d\tau}}.
\end{equation}
The two sides of Equation \eqref{eq:1.5} represent two different viewpoints on the same trajectory. On the left hand side, the trajectory is viewed as a curve (not a geodesic) in Minkowski spacetime. On the right hand side, the trajectory is seen as a geodesic  in the spacetime $g_{\alpha\beta}$.  Changing the metric $g_{\alpha\beta}$ far from the trajectory $X^{\mu}(\tau)$ does not alter the properties of the trajectory. Thus, there exist infinitely many types of $g_{\alpha\beta}$ that satisfy Equation \eqref{eq:1.5}.

A simple example is that when calculating the length of a circle defined by $(X^{1})^{2}+(X^{2})^{2}=1$, one can view the circle as an ordinary curve in the Euclidean space $ds^{2}=(dX^{1})^{2}+(dX^{2})^{2}$, and use the formula $2\int_{-1}^{1}\sqrt{1+(dX^{2}/dX^{1})^{2}}dX^{1}$ to calculate its length. One can also regard the circle as a geodesic on the two-dimensional sphere surface space $ds^{2}=g_{ij}dx^{i}dx^{j}=d\theta^{2}+\mathrm{sin}^{2}\theta d\varphi^2$, and then use the formula $\int_{\tau_{i}}^{\tau_{f}} d\tau\sqrt{g_{ij}\frac{dx^{i}}{d\tau}\frac{dx^{j}}{d\tau}}$ to calculate its length. Thus, one can obtain the equation
\begin{equation}
\label{eq:1.6}% 6
2\int_{-1}^{1}\sqrt{1+\big(\frac{dX^{2}}{dX^{1}}\big)^{2}}dX^{1}=\int_{\tau_{i}}^{\tau_{f}}d\tau\sqrt{g_{ij}\frac{dx^{i}}{d\tau}\frac{dx^{j}}{d\tau}}.
\end{equation}
The local spacetime structure far away from the circle has no influence on its properties. Thus, one can find infinite types of $g_{ij}$ that satisfy Equation \eqref{eq:1.6}.

For Minkowski spacetime, only the straight line between the initial state $|X^{\mu}(\tau_{i})\rangle$ and the final state $|X^{\mu}(\tau_{f})\rangle$ is the geodesic. In curved spacetime, it is possible that more than one trajectory in the path integral \eqref{eq:1.4} can be interpreted as the geodesics in the same spacetime. However, in this work, we consistently treat  different trajectories in the path integral \eqref{eq:1.4} as geodesics in distinct spacetimes. Each trajectory corresponds to a different spacetime. Thus, the integral over different paths in Equation \eqref{eq:1.4} can be transformed into an integral over different metrics. Therefore, using Equations \eqref{eq:1.3} and \eqref{eq:1.5}, Equation \eqref{eq:1.4} can be written as
\begin{eqnarray}\begin{split}
\label{eq:1.7} %7
\langle X^{\mu}(\tau_{f})|X^{\mu}(\tau_{i})\rangle&=N\int DX^{\mu}\ \mathrm{exp}\{i m_{0}\int_{\tau_{i}}^{\tau_{f}}d\tau\sqrt{\eta_{\alpha\beta}\frac{dX^{\alpha}}{d\tau}\frac{dX^{\beta}}{d\tau}}\}
\\&=N\int D\mathcal{F}^{\mu}\ \mathrm{exp}\{i m_{0}\int_{\tau_{i}}^{\tau_{f}}d\tau\sqrt{g_{\alpha\beta}\frac{d\mathcal{F}^{\alpha}}{d\tau}\frac{d\mathcal{F}^{\beta}}{d\tau}}\}\\&
=N\int Dg_{\mu\nu}\ \mathcal{J} \ \mathrm{exp}\{i m_{0}\int_{\tau_{i}}^{\tau_{f}}d\tau\sqrt{g_{\alpha\beta}\frac{d\mathcal{F}^{\alpha}}{d\tau}\frac{d\mathcal{F}^{\beta}}{d\tau}}\}.
\end{split}
\end{eqnarray}
Here, $\mathcal{J}$ represents the Jacobian determinant, which is induced by the transformation of the integral variables ($D\mathcal{F}^{\mu}\rightarrow Dg_{\mu\nu}$).

Note that for any trajectory $X^{\mu}(\tau)$,
there exist infinitely many spacetime metrics $g_{\alpha\beta}$ with respect to which this trajectory can be viewed as a geodesic. Spacetimes that satisfy this condition are defined to belong to the same class of spacetime, denoted by $[g_{\alpha\beta}]$, whereas spacetimes in which this trajectory cannot be regarded as a geodesic do not belong to this class. This implies that each trajectory appearing in the path integral \eqref{eq:1.4} corresponds to a distinct class of spacetimes $[g_{\alpha\beta}]$. For a specific spacetime, it belongs to, and only to, one of the classes $[g_{\alpha\beta}]$. As each trajectory is only counted once in the path integral \eqref{eq:1.4}, in each class of spacetime $[g_{\alpha\beta}]$, only one metric contributes to the path integral. However, in Equation \eqref{eq:1.7}, the integral $\int Dg_{\mu\nu}$ contains all metrics for every class of spacetime $[g_{\alpha\beta}]$. We need to eliminate the redundant metrics.

In order to eliminate redundant metrics in the path integral \eqref{eq:1.7}, only one metric from each class of spacetime $[g_{\alpha\beta}]$ should be included in the path integral. In each class of spacetime $[g_{\alpha\beta}]$, every metric is equivalent in terms of the path integral \eqref{eq:1.7}. Thus, it is permissible to select any metric from each class $[g_{\alpha\beta}]$ arbitrarily. For this purpose, one can introduce a set of gauge conditions
\begin{equation}
\label{eq:1.8}% 8
\mathcal{G}^{(m)}(g_{\alpha\beta})=0,\ \ \ \ \ \ \  \ \ m=1,\ 2,\ \ldots,\ M.
\end{equation}
Here, $M$ represents the total number of gauge conditions. Similar to the scenario with Yang--Mills gauge fields, the gauge conditions have a certain degree of arbitrariness. As long as the solution to Equation \eqref{eq:1.8} can uniquely determine a metric from each class of spacetime $[g_{\alpha\beta}]$, the gauge conditions are considered valid. For every trajectory in the path integral \eqref{eq:1.4}, there exists  a single metric that satisfies both Equations \eqref{eq:1.5} and \eqref{eq:1.8}.

According to the Faddeev--Popov method~\cite{VN,FM}, to eliminate the redundant metrics, Equation \eqref{eq:1.7} should be modified as
\begin{equation}
\label{eq:1.9}% 9
\langle X^{\mu}(\tau_{f})|X^{\mu}(\tau_{i})\rangle=N\int Dg_{\mu\nu}\ \mathcal{J} \ \Delta(\mathcal{G}^{(m)})\prod_{m=1}^{M}\delta(\mathcal{G}^{(m)})\ \mathrm{exp}\{i m_{0}\int_{\tau_{i}}^{\tau_{f}}d\tau\sqrt{g_{\alpha\beta}\frac{d\mathcal{F}^{\alpha}}{d\tau}\frac{d\mathcal{F}^{\beta}}{d\tau}}\}.
\end{equation}
Here, $\Delta(\mathcal{G}^{(m)})$ and $\delta(x)$ represent the Faddeev--Popov determinant and the $\delta$-function, respectively. The Jacobian determinant $\mathcal{J}$  is determined by Equation \eqref{eq:1.3}, while the Faddeev--Popov determinant depends on the gauge conditions \eqref{eq:1.8}. The conclusions of this work are independent of the explicit form of $\mathcal{J}$, $\Delta(\mathcal{G}^{(m)})$ and $\mathcal{F}^{\mu}$. Thus, we will avoid solving them. This is a challenging issue that needs careful study in the future.

Equation \eqref{eq:1.9} represents a new kind of representation of quantum mechanics. We name it the spacetime representation. Equation \eqref{eq:1.4} is equivalent to Equation \eqref{eq:1.9}. Thus, the integration over different trajectories can be transformed into integration over various spacetime metrics. Equation \eqref{eq:1.9} indicates that the quantum characteristics of the particle can be attributed to the uncertainty of the spacetime structure for the microscopic particle. In the conventional interpretation of path integral quantum mechanics, the spacetime structure is deterministic, and only the classical trajectory corresponds to the Hamilton--Jacobi equation. All other trajectories are not solutions of any dynamical equations. Thus, the classical trajectory  possesses a special status. However, we argued that for the free particle,  all trajectories can be seen as the geodesics in different spacetimes. Thus, different trajectories are solutions of the geodesic equation in different spacetimes. This perspective offers a more natural understanding of these trajectories. In Table~\ref{tab:1}, we outlined the key distinctions between the conventional path integral and the spacetime representation for a single relativistic free particle.
When $m_{0}\rightarrow \infty$, the microscopic particle becomes a macroscopic object. The coherence disappears and the dynamics of the particle can be described by classical physics. Consequently, the spacetime structure becomes deterministic for macroscopic objects.

\begin{table}[tbp]
\renewcommand\arraystretch{1.5}
\centering
\begin{tabular}{|l|c|c|}
\Xhline{2pt}
 Item & Conventional path integral  & Spacetime representation \\

\Xhline{0.5pt}
\makecell{Partition \\ function} & Equation \eqref{eq:1.4}

 &  Equation \eqref{eq:1.9}  \\
\Xhline{0.5pt}
\makecell{Integral\\ variable} & Particle's trajectory

 &  Spacetime metric  \\
\Xhline{0.5pt}
\makecell{Spacetime \\structure} & \parbox{6cm}{ Spacetime structure is deterministic (Minkowski spacetime).} & \parbox{6cm}{ Spacetime structure is uncertain for the microscopic particle.}\\
\Xhline{0.5pt}
\makecell{Particle's \\trajectories} &  \parbox{6cm}{ Only the classical trajectory is the solution of the Hamilton-Jacobi equation. All other trajectories in the path integral \eqref{eq:1.4} do not correspond to any dynamical equation.}  & \parbox{6cm}{ All trajectories in the path integral \eqref{eq:1.4} are viewed as geodesics in different spacetimes.}\\
\Xhline{2pt}
\end{tabular}
\caption{\label{tab:1} Two different representations of the path integral for a single relativistic free particle.}
\end{table}

\subsection{The General System}
\label{sec:1.2}

In the last subsection, we show that for a free particle, wave--particle duality can be interpreted as the uncertainty of the spacetime structure for the microscopic particle. It is well known that  the  non-commutative relationship is one of the most fundamental equations in quantum mechanics. Both wave--particle duality and the non-commutative relationship hold true regardless of whether  the particle is involved in interactions. In this subsection, we will show that in all scenarios, wave--particle duality can always be interpreted as the uncertainty of the spacetime structure for the particle. Furthermore, the path integral for any quantum system can be transformed into an integration over various spacetime metrics.

Usually, a physical system is composed of particles.  In the case where the particle number is conserved, the action of the total system can be defined as
\begin{equation}
\label{eq:1.10}% 10
S_{tot}=\int_{\sigma_{i}}^{\sigma_{f}}d\sigma\big\{\sum_{n=1}^{N} m_{n}\sqrt{g_{\alpha\beta}\frac{dX^{\alpha}_{n}}{d\sigma}\frac{dX^{\beta}_{n}}{d\sigma}}+ V(X^{\mu}_{1}, \ldots, X^{\mu}_{N})\big\}.
\end{equation}
Here,  $N$ represents the total number of the particles. $\sigma$ is an arbitrary parameter. { $X^{\mu}_{n}$ and $-m_{n}$ represent the coordinate and the rest mass of the $n$-th particle, respectively.} Furthermore, $V(X^{\mu}_{1}, \ldots, X^{\mu}_{N})$ represents the interactions between different particles.  Taking $\delta S_{tot}/\delta X^{\mu}_{n}=0$, one can obtain the classical dynamical equation of the system under general spacetime $g_{\alpha\beta}$~as
\begin{equation}
\label{eq:1.11}% 11
m_{n}\frac{d^{2}X^{\mu}_{n}}{d\sigma^{2}}+m_{n}\Gamma^{\mu}_{\alpha\beta}\frac{dX^{\alpha}_{n}}{d\sigma}\frac{dX^{\beta}_{n}}{d\sigma}+\frac{\partial V}{\partial X^{\mu}_{n}}=-m_{n}\frac{\ddot{f}_{n}}{\dot{f}_{n}}\frac{dX^{\mu}_{n}}{d\sigma}.
\end{equation}
Here, $f_{n}$ is the function that maps the proper time $\tau_{n}$ of the $n$-th particle to the parameter $\sigma$~\cite{MB13}.  It is evident that when $\partial V/\partial X^{\mu}_{n}=0$, Equation \eqref{eq:1.11} becomes the geodesic equation. Generally, when  $\partial V/\partial X^{\mu}_{n}\neq 0$, the trajectory of each particle is not the geodesic of the spacetime $g_{\alpha\beta}$.

Specifying the spacetime metric and the initial state of the system, Equation \eqref{eq:1.11} has only one solution. This solution represents the trajectories of the particles. We denote this solution as $\{X^{\mu}_{n}(\sigma)=\mathcal{K}^{\mu}_{n}(g_{\alpha\beta})\ |\ n=1,2,\ldots,N\}$. Similar to the last subsection, intuitively speaking,  changing the spacetime metric can lead the trajectories $\{X^{\mu}_{n}\ |\ n=1,2,\ldots,N\}$ to become any set of curves (physically allowed). Thus, it is reasonable to deduce that by providing any set of curves (composed of $N$ physically allowed trajectories), one can find a class of metrics $[g_{\alpha\beta}]$ in which this set of curves serves as the solution to Equation \eqref{eq:1.11}.

According to conventional quantum mechanics in Minkowski spacetime ($g_{\alpha\beta}=\eta_{\alpha\beta}$), the transition amplitude of the system defined by Equation \eqref{eq:1.10} from the initial state $| X^{\mu}_{1}(\sigma_{i}),\ldots,X^{\mu}_{N}(\sigma_{i})\rangle$ to the final state $| X^{\mu}_{1}(\sigma_{f}),\ldots,X^{\mu}_{N}(\sigma_{f})\rangle$ is
\begin{eqnarray}\begin{split}
\label{eq:1.12} %12
&\ \ \ \langle X^{\mu}_{N}(\sigma_{f}),\ldots, X^{\mu}_{1}(\sigma_{f})|X^{\mu}_{1}(\sigma_{i}),\ldots,X^{\mu}_{N}(\sigma_{i})\rangle\\&= N \int\prod_{n=1}^{N} DX^{\mu}_{n}\ \mathrm{exp}\Big\{i \int_{\sigma_{i}}^{\sigma_{f}}d\sigma\big\{\sum_{n=1}^{N} m_{n}\sqrt{\eta_{\alpha\beta}\frac{dX^{\alpha}_{n}}{d\sigma}\frac{dX^{\beta}_{n}}{d\sigma}}+ V(X^{\mu}_{1}, \ldots, X^{\mu}_{N})\big\}\Big\}.
\end{split}
\end{eqnarray}
The path integral $\int\prod_{n=1}^{N} DX^{\mu}_{n}$ represents that all possible sets of trajectories $\{X^{\mu}_{n}\ |\ n=1,2,\ldots,N\}$ are summed over.

We have argued that for any given set of trajectories $\{X^{\mu}_{n}\ |\ n=1,2,\ldots,N\}$, it is always possible to embed them into a class of metrics $[g_{\alpha\beta}]$ where the trajectories satisfy Equation \eqref{eq:1.11}. Therefore, the integration over different trajectories $\{X^{\mu}_{n}\ |\ n=1,2,\ldots,N\}$ in Equation \eqref{eq:1.12} can be transformed into an integration over various classes of spacetime metrics $[g_{\alpha\beta}]$. Following the similar statements of the last subsection, the path integral \eqref{eq:1.12} can be transformed into
\begin{eqnarray}\begin{split}
\label{eq:1.13} %13
&\ \ \ \langle X^{\mu}_{N}(\sigma_{f}),\ldots, X^{\mu}_{1}(\sigma_{f})|X^{\mu}_{1}(\sigma_{i}),\ldots,X^{\mu}_{N}(\sigma_{i})\rangle\\&= N \int Dg_{\mu\nu}\ \mathcal{J}_{N} \ \Delta(\mathcal{G}^{(m)}_{N})\prod_{m=1}^{M_{N}}\delta(\mathcal{G}^{(m)}_{N})\ \\&\ \ \ \ \ \times\mathrm{exp}\Big\{i \int_{\sigma_{i}}^{\sigma_{f}}d\sigma\big\{\sum_{n=1}^{N} m_{n}\sqrt{g_{\alpha\beta}\frac{d\mathcal{K}^{\alpha}_{n}}{d\sigma}\frac{d\mathcal{K}^{\beta}_{n}}{d\sigma}}+ V(\mathcal{K}^{\mu}_{1}, \ldots, \mathcal{K}^{\mu}_{N})\big\}\Big\}.
\end{split}
\end{eqnarray}
Here, $ \mathcal{J}_{N}$, $\Delta(\mathcal{G}^{(m)}_{N})$ and $\mathcal{G}^{(m)}_{N}$ represent the Jacobian determinant, the Faddeev--Popov determinant and the gauge conditions, respectively. $M_{N}$ represents the number of gauge conditions. These quantities differ from those of a free particle. Once again, we refrain from explicitly determining their forms in this study.

Equation \eqref{eq:1.13} is equivalent to Equation \eqref{eq:1.12}. Equation \eqref{eq:1.13} indicates that in a general system, integrating over different trajectories can be transformed into integrating over various  spacetime metrics $g_{\alpha\beta}$. In Equation \eqref{eq:1.13}, the spacetime metric is the only integral variable. Thus, wave--particle duality can generally be interpreted as the uncertainty of the spacetime structure for microscopic particles. Therefore, it is reasonable to infer that the quantum characteristics of a general system can be ascribed to the uncertainty of the spacetime structure. Moreover, this perspective implies a potential connection between quantum characteristics and geometric quantities. 

%It is worth noting that the Ryu-Takayanagi (RT) formula indicates that the entanglement entropy of a conformal field theory is equal to the area of a minimal surface~\cite{ST,VMT}.  The RT formula thus serves as a specific example illustrating the relationship between quantum characteristics and geometric quantities.

It is well known that the most successful theory (up to now) for describing interactions between microscopic particles is quantum field theory. This theory  can describe the quantum dynamics of a system whether the particle number is conserved or not. The quantum dynamics of the system defined by Equation \eqref{eq:1.10} can also be described by quantum field theory.
Thus, the equivalence between Equations \eqref{eq:1.12} and \eqref{eq:1.13} implies that the path integral over different quantum fields  can also be transformed into integration over various spacetime metrics. 

In addition, we note  that  in the limit $m_{0}\rightarrow 0$, the action in Equation \eqref{eq:1.1} vanishes. Consequently,  Equations \eqref{eq:1.9} and \eqref{eq:1.13} cannot  describe the quantum dynamics of a massless particle. The photon is an important massless particle in quantum mechanics. The self-interference of photons gives rise to the well-known double-slit interference phenomenon, and biphoton states play a significant role in quantum communications and quantum imaging~\cite{DN29}. It is interesting to note that in~\cite{DN29}, the authors showed that  biphoton states can exhibit   characteristics reminiscent  of the Yin--Yang symbol.  The photon is the quantum of the electromagnetic field. To accurately  describe the properties of photons, one must use quantum electrodynamics (QED). However, a rigorous formulation of QED in the spacetime representation is highly nontrivial and requires further careful study. By neglecting the spin degree of freedom, the photon can be approximately described by a scalar field. Discussions on formulating scalar field theory in the spacetime representation are provided in Appendix~\ref{sec:C}.

\section{Double-Slit Interference Experiment}
\label{sec:B}

{The transition amplitude in Equation  \eqref{eq:1.13} is a matrix element of the time evolution operator. Once the time evolution operator is known, it can, in principle, be used to study various quantum phenomena, such as diffraction, interference, wave packet spreading, and quantum tunneling. The double-slit interference experiment (DSIE) plays an important role in quantum mechanics. Moreover, when analyzing the DSIE interference fringes, the variation in the modulus of the transition amplitude during particle propagation can often be approximately neglected~\cite{ZZ28,EH28}, which simplifies the calculations. Therefore, as a specific example, we provide an interpretation of the DSIE from the perspective presented in this work. More complicated phenomena will require careful examination in future~studies. }

\begin{figure}[tbp]
\centering
\includegraphics[width=6cm]{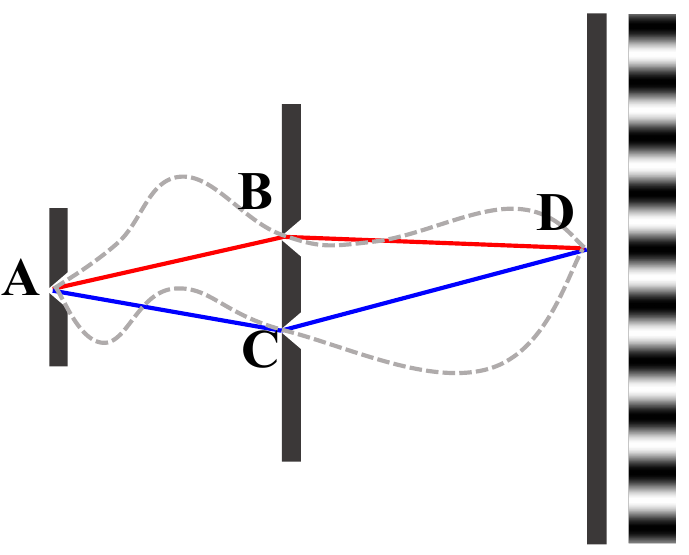}
\caption{\label{fig:m1} {Schematic diagram of the  double-slit interference experiment. Point A denotes a reference point on the incident wavefront. Points B and C represent the two slits in the double-slit apparatus. Point D is a specific observation location on the receiving screen. The interference fringes displayed on the screen, with the contrast between light and dark regions, intuitively reflect the relative strength of the particle's probability density distribution at that location.}}
\end{figure}

{  In the DSIE,  a particle interferes with itself. When the particle propagates  through spacetime, it is typically assumed that it is not subject to any interactions. Thus, we use Equation \eqref{eq:1.9} to study the resulting interference fringes.   The DSIE is illustrated in Figure~\ref{fig:m1}.   The properties of the interference fringes are governed by the transition amplitude  from  point $A$ to  point $D$, denoted by $\langle D|A\rangle$.  When calculating the interference fringes in the DSIE, variations in the modulus of the transition amplitude during particle propagation are often approximately neglected~\cite{ZZ28,EH28}. Consequently, the characteristics of the interference fringes are primarily determined by variations in the phase. For simplicity, we neglect the width of the slits. Under these approximations, the transition amplitude  $\langle D|A\rangle$ can be written as 
\begin{equation}
\label{eq:m1}
\langle D|A\rangle \propto \langle D|B\rangle \langle B|A\rangle+\langle D|C\rangle \langle C|A\rangle.
\end{equation}
Here, the definitions of $\langle B|A\rangle$, $\langle C|A\rangle$, and similar terms are analogous to that of $\langle D|A\rangle$. Equation \eqref{eq:m1} reflects the fact that, on the double-slit screen, there are only two slits through which the particle can pass. This expression is a commonly used method for calculating interference fringes.}

{ We first focus on the calculation of  $\langle B|A\rangle$. It is not necessary  to consider variations in the  modulus of the transition amplitude.  Therefore, the variations in  the prefactor $N \mathcal{J} \ \Delta(\mathcal{G}^{(m)})\prod_{m=1}^{M}\delta(\mathcal{G}^{(m)})$   in Equation \eqref{eq:1.9} can be neglected when studing the interference fringes. This implies that the prefactor can be replaced by a constant. Accordingly, based on Equation \eqref{eq:1.9}, the transition amplitude $\langle B|A\rangle$  can be written as
\begin{equation}
\label{eq:m2}
\langle B|A\rangle\propto\int Dg_{\mu\nu}\ \mathrm{exp}\{i m_{0}\int_{\tau_{A}}^{\tau_{B}}d\tau\sqrt{g_{\alpha\beta}\frac{d\mathcal{F}^{\alpha}}{d\tau}\frac{d\mathcal{F}^{\beta}}{d\tau}}\}.
\end{equation}
Here, we use  $\tau_{A}$, $\tau_{B}$, $\tau_{C}$, and so on  to  denote the moments at which the particle arrives at points A , B, C, etc., as it moves along the corresponding geodesics.  The integration  over the metric indicates that contributions from different spacetime structures must be taken into account. }

{In different spacetime configurations, different geodesics contribute to $\langle B|A\rangle$, and these geodesics generally correspond to different path lengths.  The integration over the spacetime metric in Equation \eqref{eq:m2} is difficult to calculate. For simplicity, we adopt saddle-point approximation, which typically works well in the semiclassical region. In this limit, only the contribution from the spacetime configuration satisfying  $\delta S/\delta g_{\mu\nu}=0$ is taken into account. Consequently,  Equation \eqref{eq:m2} reduces to 
\begin{equation}
\label{eq:m3}
\langle B|A\rangle \propto\ \mathrm{exp}\{i m_{0}\int_{\tau_{A}}^{\tau_{B}}d\tau\sqrt{g^{min}_{\alpha\beta}\frac{d\mathcal{F}^{\alpha}}{d\tau}\frac{d\mathcal{F}^{\beta}}{d\tau}}\}.
\end{equation}
Here,  $g^{min}_{\alpha\beta}$ denotes the spacetime metric  corresponding to the minimal action of the particle. Since the particle action is proportional to the length of its trajectory, the minimal action corresponds to the shortest path.  The shortest trajectory between points $A$ and $B$ is a straight line, as shown in Figure~\ref{fig:m1}. This straight line is  a geodesic in Minkowski spacetime. Therefore, 
 under the saddle point approximation,  the metric $g^{min}_{\alpha\beta}$ should be taken as  the Minkowski  metric, i.e.,  $g^{min}_{\alpha\beta}=\eta_{\alpha\beta}$.  }

{In  Minkowski spacetime, the Lagrangian of the particle is
\begin{equation}
\label{eq:m4}
L=m_{0}\sqrt{g^{min}_{\alpha\beta}\frac{d\mathcal{F}^{\alpha}}{d\tau}\frac{d\mathcal{F}^{\beta}}{d\tau}}=  m_{0}\sqrt{\eta_{\alpha\beta}\frac{d\mathcal{F}^{\alpha}}{d\tau}\frac{d\mathcal{F}^{\beta}}{d\tau}}.
\end{equation}
Taking the nonrelativistic limit, the Lagrangian $L$ becomes 
 \begin{equation}
\label{eq:m5}
L\approx m_{0}-\frac{1}{2}m_{0}\eta_{ii}\frac{d\mathcal{F}^{i}}{d\tau}\frac{d\mathcal{F}^{i}}{d\tau}.
\end{equation}
Here,~$\eta_{ii}$ and $\mathcal{F}^{i}$ denote the spatial components of the corresponding quantities. Since $\mathcal{F}^{\beta}$ represents the particle trajectory,  $\frac{d\mathcal{F}^{i}}{d\tau}$ corresponds to the $i-th$ component of the velocity. For convenience, we denote it by $v^{i}$. Consequently, Equation \eqref{eq:m5} can be written as
 \begin{equation}
\label{eq:m6}
L\approx m_{0}-\frac{1}{2}m_{0}\vec{v}^{2},
\end{equation}
where  $\vec{v}^{2}\equiv (v^{1})^{2}+(v^{2})^{2}+(v^{3})^{2}$. Since we use $-m_{0}$ to denote the rest mass of the particle, the second term on the right hand side of Equation \eqref{eq:m6} represents the kinetic energy, while the first term $m_{0}$ is a constant and does not affect the dynamics. }

{For the straight-line trajectory between points  $A$ and $B$, the velocity  $\vec{v}$ can be expressed~as
\begin{equation}
\label{eq:m7}
\vec{v}=\frac{X_{B}-X_{A}}{\tau_{B}-\tau_{A}}.
\end{equation}
Here, $X_{A}$,  $X_{B}$,  $X_{C}$, and so on denote the spatial coordinates of points $A$, $B$, $C$, etc., respectively.
Substituting Equations \eqref{eq:m6} and \eqref{eq:m7} into Equation \eqref{eq:m3}, the transition amplitude $\langle B|A\rangle$ becomes
\begin{equation}
\label{eq:m8}
\langle B|A\rangle\propto\ \mathrm{exp}\Big\{i\int_{\tau_{A}}^{\tau_{B}}d\tau\Big(m_{0}-\frac{1}{2}m_{0}(\frac{X_{B}-X_{A}}{\tau_{B}-\tau_{A}})^{2}\Big) \Big\}.
\end{equation}
Following similar derivations, one can show that the transition amplitudes $\langle C|A\rangle$, $\langle D|B\rangle$, and $\langle D|C\rangle$ take forms analogous to that in Equation \eqref{eq:m8}.  Substituting these amplitudes into Equation \eqref{eq:m1}, one finally obtains the transition amplitude $\langle D|A\rangle$ as
\begin{eqnarray}\begin{split}
\label{eq:m9}  
\langle D|A\rangle \propto \ & \mathrm{exp}\{im_{0}(\tau_{D}-\tau_{A})\}\mathrm{exp}\Big\{-\frac{i}{2}m_{0}\Big(\frac{(X_{D}-X_{B})^{2}}{\tau_{D}-\tau_{B}}+\frac{(X_{B}-X_{A})^{2}}{\tau_{B}-\tau_{A}}\Big)\Big\}\\& + \mathrm{exp}\{im_{0}(\tau_{D}-\tau_{A})\}\mathrm{exp}\Big\{-\frac{i}{2}m_{0}\Big(\frac{(X_{D}-X_{C})^{2}}{\tau_{D}-\tau_{C}}+\frac{(X_{C}-X_{A})^{2}}{\tau_{C}-\tau_{A}}\Big)\Big\}.
\end{split}
\end{eqnarray}
Equation \eqref{eq:m9} indicates that under saddle-point approximation, the dominant contributions to the transition amplitude $\langle D|A\rangle$ arise from the  red and blue trajectories  depicted in~Figure~\ref{fig:m1}.}

{Assuming that the particle's speed $|\vec{v}|$ remains constant while passing through the double-slit screen, and approximately neglecting the speed difference between the red and blue trajectories in Figure~\ref{fig:m1}, Equation \eqref{eq:m9} can be simplified using the relationship between speed, distance, and time. The simplified expression becomes 
\begin{eqnarray}\begin{split}
\label{eq:m10}  
\langle D|A\rangle \propto \ & \mathrm{exp}\{im_{0}(\tau_{D}-\tau_{A})\}\mathrm{exp}\Big\{-\frac{i}{2}m_{0}|\vec{v}|(\ell_{DB}+\ell_{BA})\Big\}\\& + \mathrm{exp}\{im_{0}(\tau_{D}-\tau_{A})\}\mathrm{exp}\Big\{-\frac{i}{2}m_{0}|\vec{v}|(\ell_{DC}+\ell_{CA})\Big\}.
\end{split}
\end{eqnarray}
Here,  $\ell_{BA}$ represents the distance (defined in Minkowski spacetime) between  points  $A$ and $B$. The definitions of $\ell_{DB}$,  $\ell_{DC}$, and  $\ell_{CA}$ are analogous to that of $\ell_{BA}$. In Equation \eqref{eq:m10}, the term $\ell_{DB}+\ell_{BA}$ represents the length of the red  trajectory in Figure~\ref{fig:m1}, while  $\ell_{DC}+\ell_{CA}$ represents the length of the blue trajectory in Figure~\ref{fig:m1}. }

{The transition probability from  point $A$ to  point $D$ is given by $P_{AD}=|\langle D|A\rangle|^{2}$. From Equation \eqref{eq:m10}, we obtain 
\begin{equation}
\label{eq:m11}
P_{AD}\propto 2+2\ \textmd{cos}\Big(-\frac{1}{2}m_{0}|\vec{v}|(\ell_{DB}+\ell_{BA}-\ell_{DC}-\ell_{CA})\Big).
\end{equation}
Equation \eqref{eq:m11} indicates that the characteristics of the interference fringes are determined by the particle momentum and the difference in length between the red and blue trajectories shown in Figure~\ref{fig:m1}. This is a typical feature of the double-slit interference phenomenon~\cite{ERH28}.   Moreover, Equation \eqref{eq:m11} can also be derived from  conventional path integral quantum mechanics~\cite{ERH28}. Therefore, the DSIE can be interpreted from the perspective presented in this work.}

\section{A Proposal for Quantum Gravity}
\label{sec:2}
Although  a self-consistent and complete quantum gravity theory is still lacking, there have been some exciting attempts to construct the quantum gravity theory, such as string theory, holographic duality,  loop quantum gravity,  and more~\cite{CK,CR,JM,EW,OSJ}. Perhaps the most conservative approach is path integral quantum gravity,
\begin{equation}
\label{eq:2.1}% 17
\mathcal{Z}=\int Dg_{\mu\nu}D\varphi\ \mathrm{exp}\{i S(g_{\mu\nu},\varphi)\}
\end{equation}
or canonical quantum gravity,
\begin{equation}
\label{eq:2.2}% 18
\hat{H}(g_{\mu\nu},\varphi)|\Psi(g_{\mu\nu},\varphi) \rangle=0.
\end{equation}
In Equation \eqref{eq:2.1}, $\varphi$ represents the  matter fields, and $S(g_{\mu\nu},\varphi)$ represents the total action of the system. Equation \eqref{eq:2.2} is the
Wheeler--De Witt equation. It is widely believed that the path integral approach \eqref{eq:2.1} and the canonical quantization approach should be equivalent if the Hamiltonian operator $\hat{H}$ is properly defined. We will focus on these two approaches of  quantum gravity.

There are some divergence difficulties that have not been fully resolved in the path integral approach~\cite{GW,GW2}. The time problem in the Wheeler--De Witt equation has also not been fully addressed~\cite{CK,HJ,KV,CJ}. The aim of this study is not to solve all these well-known problems. Instead, we will demonstrate that there may be another issue present in both conventional path integral quantum gravity and  canonical quantum gravity. We also propose a solution to circumvent this issue based on our discussions in Section~\ref{sec:1}.

Consider a simple example of a flat ($k=1$) FRW  universe. In conformal time coordinates, the classical spacetime metric is
\begin{equation}
\label{eq:2.3}% 19
ds^{2}=a^{2}(dt^{2}-dx^{2}-dy^{2}-dz^{2}).
\end{equation}
Here, $a$ represents the scale factor.  If there is a massless real scalar field minimally coupled to the FRW spacetime, the action of the total system is~\cite{HJ}
\begin{equation}
\label{eq:2.4}% 20
S(g_{\mu\nu}, \phi)=\frac{1}{16\pi}\int d^{4}x\sqrt{-g}R+ \frac{1}{2}\int d^{4}x\sqrt{-g}g^{\mu\nu}\partial_{\mu}\phi\partial_{\nu}\phi +S_{\partial M}.
\end{equation}
On the right hand side of Equation \eqref{eq:2.4}, the first term is the Einstein--Hilbert action. The second term represents the action of the scalar field.  $R$  and $S_{\partial M}$ represent the Ricci scalar and the Gibbons–Hawking–York surface term, respectively.

Combining Equations \eqref{eq:2.3} and \eqref{eq:2.4}, one can prove that the total Hamiltonian operator of the system described by \eqref{eq:2.4} can be approximately written as~\cite{HJ}
\begin{equation}
\label{eq:2.5}% 21
\hat{H}_{tot}=-\frac{2\pi}{3V_{0}}\hat{\pi}_{a}^{2}+\frac{V_{0}}{(2\pi)^{3}}\int d\vec{k}^{3}|\vec{k}|A_{\vec{k}}^{\dag}A_{\vec{k}}.
\end{equation}
Here, $V_{0}$ and $\hat{\pi}_{a}$ represent the coordinate volume and the conjugate momentum operator, respectively. $A_{\vec{k}}^{\dag}$ ($A_{\vec{k}}$) represents the creation (annihilation) operator of the scalar particle with momentum $\vec{k}$. In Equation \eqref{eq:2.5}, the vacuum energy of the scalar field has been neglected.
The conjugate momentum $\pi_{a}$ is proportional to the coordinate volume $V_{0}$. Thus, one can prove that the coordinate volume $V_{0}$ is a trivial global conformal factor, which can be eliminated from the theory~\cite{HJ}. Therefore, it does not affect the result. 

%The Hamiltonian operator \eqref{eq:2.5} implies that the scalar particle number is conserved. This is reasonable, especially in the semiclassical region,  since massless scalar particles described by Equation \eqref{eq:2.4} can not be produced from the %vacuum state in the FRW spacetime~\cite{SY}. In the case where the scalar field is in a thermal state and the total energy of all scalar particles is significantly higher than the vacuum energy, one can disregard the vacuum energy of the scalar field.

Since the canonical  Hamiltonian of the total system is equal to zero, one can not obtain the dynamical information of the total universe from the Wheeler--De Witt equation. However, the (effective) Hamiltonian of a subsystem is often non-zero~\cite{HJ2}. This does not violate general covariance. The quantum dynamics of the subsystem can be described by the  von  Neumann Equation \cite{HJ,HJ2,HJ3,HF}
\begin{equation}
\label{eq:2.6}% 22
\frac{d\rho}{dt}=-i \mathrm{Tr}_{B}[\hat{H}_{tot}, \rho_{tot}].
\end{equation}
Here, $\rho$ and $t$ represent the reduced density matrix of the subsystem and the coordinate time variable, respectively. $\mathrm{Tr}_{B}$ represents the partial trace over the environment,
and $\rho_{tot}$ represents the density matrix of the total system. The variable $t$ can also be interpreted as a Brown--Kucha$\check{\mathrm{r}}$ dust field~\cite{HJ,HJ2}. In fact, Equation \eqref{eq:2.6} can be derived from the Wheeler--De Witt equation by introducing the Brown--Kucha$\check{\mathrm{r}}$ dust field~\cite{BK,VT,HJ,HJ2}. It should be noted that when using Equation \eqref{eq:2.6} to describe the dynamics of the subsystem, one must impose the constraint $\mathrm{Tr}(\hat{H}_{tot}\rho_{tot}(t_{i}))=0$~\cite{HJ,HJ2}, where $\rho_{tot}(t_{i})$ represents the initial density matrix of the universe.

%Based on Equation \eqref{eq:2.6},  we studied the evolution of radiation, non-relativistic matters or dark energy dominated quantum universe in~\cite{HJ}. In all these cases, we show that the classical trajectory of the universe is consistent with the %quantum evolution of the wave packet. One can obtain this result in different coordinates. In 2015, Maeda employed a different approach to obtain a similar result for the dark energy dominated universe~\cite{HM}.

One can use the Hamiltonian operator \eqref{eq:2.5} to describe the heat radiation dominated universe. Substituting Equation \eqref{eq:2.5} into Equation \eqref{eq:2.6}, we find that even if the heat radiation is in a classical state,  the coherence of spacetime steadily increases with time~\cite{HJ}. This indicates that the quantum characteristics of the spacetime are becoming increasingly significant. The same result can be derived using either  conformal time coordinates or  proper time coordinates. For the  non-relativistic classical matter-dominated universe, we also arrive at the same conclusion~\cite{HJ}. However, we believe that this result is not quite reasonable. When the matter field is in a classical state, it can be described by a classical energy-momentum tensor. General relativity implies that the spacetime structure is determined by the energy-momentum tensor of the matter field. Then, it is natural to expect that a classical energy-momentum tensor should correspond to a classical spacetime structure.  Therefore, there exists an issue in conventional canonical quantum gravity. Specifically, classical matter fields may be associated with a quantum spacetime. This issue also appears in conventional path integral  quantum gravity~\cite{HJ}.

According to Equations \eqref{eq:2.1} and \eqref{eq:2.2}, in order to quantize a gravitational system, it is necessary to simultaneously quantize the spacetime and all matter fields. In the path integral \eqref{eq:2.1}, the spacetime metric and the matter fields are independent integral variables. This leads to the issue that  classical matter may correspond to a quantum spacetime. However, according to our statements in Section~\ref{sec:1}, wave--particle duality can be interpreted as the uncertainty of the spacetime structure for the microscopic particle. The quantum characteristics of any system can be attributed to the uncertainty of the spacetime structure. In other words, the quantization of any system can be achieved by integrating over various spacetime metrics.  We believe this viewpoint is also valid for gravitational systems. Therefore, when quantizing a gravitational system via the path integral approach, one should take only the spacetime metric as the integral variable.

Taking these considerations into account, we propose that the conventional path integral quantum gravity \eqref{eq:2.1} should be modified as
\begin{equation}
\label{eq:2.7}% 23
\mathcal{Z}=\int Dg_{\mu\nu}\  \mathrm{exp}\{i S(g_{\mu\nu}, \varphi=\Phi(g_{\mu\nu}))\}.
\end{equation}
Here, $\varphi=\Phi(g_{\mu\nu})$ represents the solution to the classical dynamical equation
\begin{equation}
\label{eq:2.8}% 24
\frac{\delta S(g_{\mu\nu}, \varphi)}{\delta \varphi}=0.
\end{equation}
Specifying the initial state of the matter fields and the spacetime metric, the dynamical Equation \eqref{eq:2.8} yields a unique solution. Thus, $\varphi=\Phi(g_{\mu\nu})$ exists. Certain factors (such as  the Faddeev--Popov determinant and the $\delta$-function) may need to be included in \mbox{Equations~\eqref{eq:2.7} and \eqref{eq:2.1}} to remove gauge degrees of freedom. These factors depend on the matter fields. We refrain from delving into this issue in the present study.

In conventional path integral quantum gravity \eqref{eq:2.1}, both the spacetime metric and the matter fields serve as integral variables. The matter field configurations do not necessarily have to be solutions of the Hamilton--Jacobi equation. However, in Equation \eqref{eq:2.7}, only the spacetime metric serves as the integral variable.  In Section~\ref{sec:1}, we have shown that when  the quantum characteristics of the system are attributed to the uncertainty of the spacetime structure, we must consider different matter configurations as solutions of the Hamilton--Jacobi equation in various spacetimes.  Therefore, in Equation \eqref{eq:2.7}, we set $\varphi=\Phi(g_{\mu\nu})$.

In the Wheeler--De Witt equation, Equation \eqref{eq:2.2}, the spacetime metric $g_{\mu\nu}$ and the matter fields $\varphi$ represent independent degrees of freedom. However, according to our statements, in the canonical Hamiltonian operator  of the total system, the degrees of freedom should only include the spacetime metric $g_{\mu\nu}$. Analogous to the case of the path integral quantum gravity, one can use the equation $\varphi=\Phi(g_{\mu\nu})$ to eliminate the degrees of freedom $\varphi$ in the Wheeler--De Witt equation.

First, by substituting the equation $\varphi=\Phi(g_{\mu\nu})$ into the classical canonical Hamiltonian $H(g_{\mu\nu}, \varphi)$ of the total system, the Hamiltonian transforms into $\mathbb{H}(g_{\mu\nu})=H(g_{\mu\nu}, \Phi(g_{\mu\nu}))=0$. Next, upon quantizing the total system, we have $\mathbb{H}(g_{\mu\nu})\rightarrow \hat{\mathbb{H}}(g_{\mu\nu})$. Consequently, one can obtain the equation
\begin{equation}
\label{eq:2.9}% 25
\hat{\mathbb{H}}(g_{\mu\nu})|\Psi(g_{\mu\nu}) \rangle=0.
\end{equation}
In principle, by solving Equation \eqref{eq:2.9}, one can determine the distribution of various spacetime metrics, provided that the inner product of the Hilbert space is well defined.  In each possible spacetime $g_{\mu\nu}$, the dynamics of matter are described by $\varphi=\Phi(g_{\mu\nu})$. The superposition of the dynamics across different spacetimes yields the quantum dynamics of the matter fields. This is akin to the scenario of a single particle, as discussed in Section~\ref{sec:1.1}.

The philosophy hidden behind Equations \eqref{eq:2.7} and \eqref{eq:2.9} is that the quantum characteristics of the matter fields are attributed to the uncertainty of the spacetime structure. When matter fields are in a quantum state, this implies that  the spacetime metric is uncertain. In other words, the spacetime  is also in a quantum state.  The quantum spacetime state can be generally written as $\sum_{n}c_{n}|g^{(n)}_{\mu\nu}\rangle$, where $\{|g^{(n)}_{\mu\nu} \rangle \ |\ n=1, 2, \ldots\}$ represents a set of orthonormalized and complete eigenstates of the metric operator $\hat{g}_{\mu\nu}$. Conversely, when matter fields are in a classical state,  this implies that the metric is definite, and the spacetime is in a classical state. Therefore, in Equations \eqref{eq:2.7} and \eqref{eq:2.9}, there is no issue that classical matter fields correspond to a quantum spacetime.

%{To sum up, we show that the quantum characteristics for a general system can be ascribed to the uncertainty of the spacetime structure for the microscopic particles and the quantization can  be achieved by integration over various spacetime metrics. We believe that this viewpoint should also be valid for the gravitational system. According to this viewpoint, we introduce a proposal that when quantizing gravitational system via the path integral approach, the spacetime metric should be served as the unique integral variable. This proposal leads to a different quantum gravity model. In this model, the issue appeared in the conventional quantum gravity model that classical matter fields may correspond to quantum spacetime naturally disappear. Therefore, our result may help to the study of the quantum gravity.}

{Furthermore, in Section~\ref{sec:1}, we pointed out that Equations \eqref{eq:1.12} and \eqref{eq:1.13} are equivalent. Thus, in systems without gravitational effects, it seems difficult to experimentally distinguish the viewpoint presented in this work from the conventional interpretation of quantum mechanics. However, for gravitational systems, our viewpoint naturally leads to a distinct  model of quantum gravity. With advances in experimental techniques, it may become possible to test different quantum gravity models experimentally in the future. This implies that, in principle, the model proposed in this section can be examined by future experiments. Such tests may provide a way to experimentally evaluate the interpretation of quantum mechanics proposed in this work. For example, consider a given isolated region (where both the matter and the spacetime structure of this region are not impacted by other regions), and assume  all matter $\varphi$, including the experimental apparatus, is in a classical state. Our analysis in this section indicates that an experiment would then find that the spacetime structure in this region must also be classical. In other words, the experiment would  detect a definite  spacetime structure in this region. Therefore, if experiments were to show that the spacetime structure in such a region cannot be in a superposition state $\sum_{n}c_{n}|g^{(n)}_{\mu\nu}\rangle$, this would support the results of this work. }

\section{Conclusions and Discussions}
\label{sec:3}

In conventional path integral quantum mechanics, one needs to integrate over all possible trajectories in the phase space of the system. We demonstrate that this integration can be transformed into integrating over all spacetime metrics, leading to the spacetime representation of the quantum mechanics.
The path integral quantum mechanics can be represented in the spacetime representation, indicating that the quantum properties of any system can be ascribed to the uncertainty of the spacetime structure for microscopic particles.  On the other hand, in the conventional interpretation of quantum mechanics, all quantum phenomena are induced by wave--particle duality. Thus, wave--particle duality can be interpreted as the uncertainty of the spacetime structure for microscopic particles. Therefore, the philosophical basis of quantum mechanics may be viewed as the uncertainty of the spacetime structure.

We show that different trajectories in conventional path integral quantum mechanics can be regarded as solutions of the Hamilton--Jacobi equation in various spacetimes. In particular, for the free particle, different trajectories in the conventional path integral can be viewed as geodesics in different spacetimes.  This perspective appears more natural than the conventional one. In conventional path integral quantum mechanics, only the classical trajectory satisfies the Hamilton--Jacobi equation. All  other trajectories do not satisfy the dynamical equation.

In conventional path integral quantum gravity, both the spacetime metric and the matter fields act as integral variables. This results in a situation where classical matter fields may correspond to quantum spacetime. We argued that this is not quite reasonable. According to our assertions, the quantum characteristics of matter are induced by the uncertainty of the spacetime structure. This implies that when using the path integral method to quantize any system,  we only need to set the spacetime metric as the unique integral variable. Therefore, we propose that in path integral quantum gravity, the matter fields should not be considered the integral variables. Based on this proposal, we present a revised version of path integral quantum gravity. In this version,  the classical matter fields always correspond to the classical spacetime.

\section*{Acknowledgements}
Hong Wang  was supported by the National Natural Science Foundation of China Grant  No. 12234019. Hong Wang thanks for the help from Professor Erkang Wang.

\appendix
\section{Viewing any one trajectory in equation \eqref{eq:1.4} as a geodesic}
\label{sec:A}

In Section~\ref{sec:1.1}, based on some arguments, we claim that any trajectory in Equation \eqref{eq:1.4} can be embedded into a  spacetime $g_{\alpha\beta}$ where it becomes a geodesic. We provide a more rigorous demonstration in the scenario where the spacetime dimension is two.

\begin{figure}[tbp]
 \centering
\includegraphics[width=12cm]{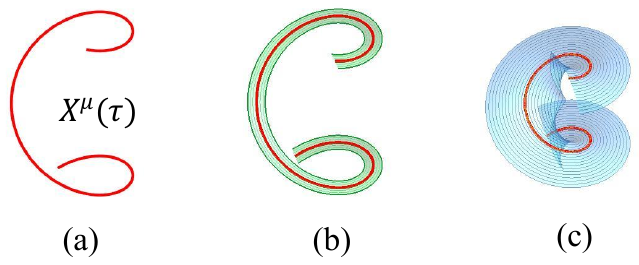}
\caption{\label{fig:1} Embedding any one trajectory in the path integral \eqref{eq:1.4} into a curved spacetime where it becomes a geodesic. The red curve in  (\textbf{a}) 
 represents an arbitrary trajectory $X^{\mu}(\tau)$ in the path integral~\eqref{eq:1.4}.   (\textbf{b}) shows an infinitely thin ribbon where the trajectory $X^{\mu}(\tau)$ acts as a geodesic.   (\textbf{c})~depicts a two-dimensional surface. The ribbon in  (\textbf{b}) is a part of this surface.}
\end{figure}

One can denote an arbitrary trajectory in Equation \eqref{eq:1.4} as $X^{\mu}(\tau)$, as shown in Figure~\ref{fig:1}a. $X^{\mu}$ and $\tau$ represent the coordinates of Minkowski spacetime and the proper time of the particle, respectively. Thus, the unit tangent vector of this trajectory at the moment $\tau$ is $\frac{dX^{\mu}(\tau)}{d\tau}$. The derivative of the vector $\frac{dX^{\mu}(\tau)}{d\tau}$ with respect to the proper time $\tau$ is~\cite{WHC}
\begin{equation}
\label{eq:A1}
\frac{d}{d\tau}\Big(\frac{dX^{\mu}(\tau)}{d\tau}\Big)=\kappa(\tau) \vec{n}(\tau).
\end{equation}
Here, $\kappa(\tau)$ and $\vec{n}(\tau)$ represent the curvature and the principle normal vector of the trajectory in the moment $\tau$, respectively.

$\epsilon(\tau)$ is defined as a two-dimensional  infinitesimal domain that satisfies the following~conditions:

\begin{enumerate}
\item[(\romannumeral1).]   $\epsilon(\tau)$ is a part of Minkowski spacetime.
\item[(\romannumeral2).]  The point $\tau$ of the trajectory $X^{\mu}(\tau)$ belongs to $\epsilon(\tau)$.
\item[(\romannumeral3).]  $\vec{n}(\tau)$ is the normal vector of $\epsilon(\tau)$.
\end{enumerate}

At any point along the trajectory $X^{\mu}(\tau)$, one can always define such an infinitesimal domain $\epsilon(\tau)$. The union of all these domains is denoted as $\cup \epsilon(\tau)$. Then $\cup \epsilon(\tau)$ is an infinitely thin ribbon to which the trajectory belongs, as shown in Figure~\ref{fig:1}b.

The dimension of the infinitely thin ribbon $\cup \epsilon(\tau)$  is two. This ribbon can be extended into a two-dimensional curved spacetime, as shown in Figure~\ref{fig:1}c. We denote the metric of this curved spacetime as $g_{\alpha\beta}$. As long as the ribbon $\cup \epsilon(\tau)$ is a part of the extended spacetime, the extension of the ribbon is valid. Thus, there are infinite ways to extend the ribbon $\cup \epsilon(\tau)$.

At any point along the trajectory $X^{\mu}(\tau)$, the principle normal vector $\vec{n}(\tau)$ of the trajectory is the normal vector of the infinitesimal domain $\epsilon(\tau)$. Thus, at any point on the trajectory, $\vec{n}(\tau)$ is also the normal vector of the extended spacetime $g_{\alpha\beta}$. In general, a curve is a geodesic on a surface if, at each point of the curve, its principal normal vector is the normal vector of the surface~\cite{WHC}.  Hence, the trajectory $X^{\mu}(\tau)$ is a geodesic in the spacetime $g_{\alpha\beta}$. Therefore, given an arbitrary trajectory in the path integral \eqref{eq:1.4}, one can always embed it into a curved spacetime where it becomes a geodesic. While our demonstrations are focused on the two-dimensional spacetime, we believe that this conclusion holds true in higher dimensional spacetime.

According to the conventional interpretation of quantum mechanics, the path integral in Equation \eqref{eq:1.4} includes some discrete or non-differentiable trajectories. One may argue that such trajectories cannot be regarded as geodesics in a smooth and continuous spacetime. However, we believe that our demonstration in this section can be extended to non-smooth or non-continuous spacetimes. This suggests that the discrete or non-differentiable trajectories in Equation \eqref{eq:1.4} can still be interpreted as geodesics in a certain class of non-smooth or non-continuous spacetimes. If these trajectories are physically allowed, then the path integral in Equation \eqref{eq:1.9} should involve contributions from such non-smooth or non-continuous spacetimes. Nevertheless, to our knowledge, the existence of these trajectories has yet to be directly verified experimentally. Furthermore, the mathematics of non-smooth or non-continuous spacetimes is complicated. For these reasons, we will not explore these trajectories in detail in the present work.

\section{Field Systems}

 \label{sec:C}

{For simplicity, we consider the case of the scalar field. }
The action of a scalar field, minimally coupled with the curved spacetime, is defined as~\cite{DB}
\begin{equation}
\label{eq:1.14}% 14
S_{\phi}=\int d^{4}x \sqrt{-g}\{\frac{1}{2}g^{\mu\nu}\partial_{\mu}\phi\partial_{\nu}\phi-V(\phi)\}.
\end{equation}
Here, $\phi$ and $V(\phi)$ represent the scalar field and the scalar potential, respectively. Taking $\delta S_{\phi}/\delta \phi=0$, one can obtain the classical dynamical equation of the scalar field as~\cite{DB}
\begin{equation}
\label{eq:1.15}% 15
\frac{1}{\sqrt{-g}}\partial_{\mu}(\sqrt{-g}\partial^{\mu}\phi)+\frac{\partial V}{\partial \phi}=0.
\end{equation}
Given the initial state of the scalar field and the spacetime metric $g_{\alpha\beta}$, Equation \eqref{eq:1.15} has a unique solution. We denote this solution as $\phi=\Phi(g_{\alpha\beta})$. By varying the metric $g_{\alpha\beta}$, the solution $\phi=\Phi(g_{\alpha\beta})$ can be transformed into any physically allowed scalar field configuration. Thus, conversely, it is reasonable to infer that given any physically allowed  scalar field configuration, one can find a spacetme metric in which this field configuration serves as the solution to Equation \eqref{eq:1.15}. Therefore, integration over different field configurations can be transformed into integration over various metrics.

Following the statements in the last subsection, the partition function of the scalar field in Minkowski spacetime
\begin{equation}
\label{eq:1.16}% 16
\mathcal{Z}=\int D\phi \ \mathrm{exp}\big\{i\int d^{4}x \{\frac{1}{2}\eta^{\mu\nu}\partial_{\mu}\phi\partial_{\nu}\phi-V(\phi)\}\big\}
\end{equation}
can be equivalently written as
\begin{equation}
\label{eq:1.17}% 16
\mathcal{Z}=\int Dg_{\mu\nu} \ \mathcal{J}_{\phi} \ \Delta(\mathcal{G}^{(m)}_{\phi})\prod_{m=1}^{M_{\phi}}\delta(\mathcal{G}^{(m)}_{\phi})\ \mathrm{exp}\big\{i\int d^{4}x \sqrt{-g}\{\frac{1}{2}g^{\alpha\beta}\partial_{\alpha}\Phi\partial_{\beta}\Phi-V(\Phi)\}\big\},
\end{equation}
where $ \mathcal{J}_{\phi}$, $\Delta(\mathcal{G}^{(m)}_{\phi})$ and $\mathcal{G}^{(m)}_{\phi}$ represent the Jacobian determinant, the Faddeev--Popov determinant and the gauge conditions, respectively. $M_{\phi}$ is the number of gauge conditions.  For the quantum field theory, it seems possible that $M_{\phi}=0$. In this case, when changing the path integral variable $D\phi\rightarrow Dg_{\mu\nu}$, one does not need to introduce any gauge conditions. However, the conclusions of this work remain unaffected by the specific value of $M_{\phi}$.

We point out that in Equation \eqref{eq:1.17}, although the path integral is performed over various spacetime metrics, the impact of the matter field on the spacetime structure (the effect of general relativity) is not taken into account. Equation \eqref{eq:1.17} solely indicates that the non-commutative relationship can be interpreted as the uncertainty of the spacetime structure for the scalar field $\phi$.  For a field with a non zero spin, there may exist a correlation between the spin and the spacetime torsion~\cite{FW,MB,MX}. We do not delve into these more complex fields in this study. However, based on our discussions,  it is reasonable to deduce that the quantum characteristics of any quantum field can be ascribed to the uncertainty of the spacetime structure.

\end{document}